\documentclass[preprint,11pt]{elsarticle}

\usepackage{graphicx}
\usepackage{multirow}
\usepackage{amssymb}
\usepackage{float}
\usepackage{amsmath}
\usepackage{xcolor}
\usepackage{natbib}
\usepackage{setspace}

\usepackage{geometry}
\usepackage{mathrsfs}
\usepackage{float}
\usepackage[multiple]{footmisc}
\usepackage[final]{changes}

\usepackage{graphicx}

\journal{}
\begin{document}

\begin{frontmatter}



\title{Clustering coefficients as measures of the complex interactions in a directed \added{weighted multilayer} network}

		\author[Bic1]{Paolo Bartesaghi}
		\author[Catt]{Gian Paolo Clemente}
		\author[Bic1]{Rosanna Grassi\corref{cor1}}
		
		\cortext[cor1]{\emph{Corresponding author. email: rosanna.grassi@unimib.it}} 
		\address[Bic1]{University of Milano - Bicocca, Via Bicocca degli Arcimboldi 8, 20126 Milano, Italy, email: paolo.bartesaghi@unimib.it; rosanna.grassi@unimib.it}
		\address[Catt]{Universit\`{a} Cattolica del Sacro Cuore di Milano, Largo Gemelli 1, 20123 Milano, Italy email: gianpaolo.clemente@unicatt.it}

\begin{abstract}
In this paper, we provide novel definitions of clustering coefficient for weighted and directed multilayer networks. We extend in the multilayer theoretical context the clustering coefficients proposed in the literature for weighted directed monoplex networks. We quantify how deeply a node is involved in a choesive structure focusing on a single node, on a single layer or on the entire system. The coefficients convey several characteristics inherent to the complex topology of the multilayer network. We test their effectiveness applying them to a particularly complex structure such as the international trade network. 
The trade data integrate different aspects and they can be described by a directed and weighted multilayer network, where each layer represents import and export relationships between countries for a given sector. The proposed coefficients find successful application in describing the interrelations of the trade network, allowing to 
disentangle the effects of countries and sectors and jointly consider the interactions between them.
\end{abstract}


\end{frontmatter}


\section{Introduction}
\label{Introduction}


The occurrence of multiple interactions within networked systems has driven the study of the multilevel nature of real-world networks. In this setting, increasing attention has been paid to heterogeneous networks in the literature. These networks have different classifications at present, including heterogeneous information networks (\cite{GUPTA2020,Chen2021}), multilayer or multiplex networks (\cite{DeDomenico2013,Kivela2014,Boccaletti2014}) and multidimensional networks (\cite{Berlingerio2013}). 
Independently of this classification, relations of a different nature are incorporated in the same complex structure.  
Indeed, networks usually present different characteristics in the various layers that cannot be adequately described by aggregating the layers in the overlay or the projected network (see, e.g., \cite{Battiston2014}). Furthermore, the meaning of the interconnections between nodes of different layers goes beyond a simple formal representation.
These aspects prompt the study of complex structures represented by multilayer networks that preserve their original architecture as much as possible.

In this field, specific interest is posed to the triangles that a node can form across the multilayer network, \deleted{especially in a multiplex network with non-diagonal couplings}. Indeed, in this case, the adjacent nodes of a node can belong to the same or to different layers (see \cite{Cozzo2015}). 
Since the occurrence of triangle patterns in the network is usually measured by the well-known clustering coefficient, different triangle definitions affect the expression of the clustering coefficient in multilayer systems. The clustering coefficient is in fact an important measure of network topology (see \cite{Newman, Watts_1998}). In the application domain, it is meant to measure the extent to which nodes in a graph tend to cluster, representing an effective measure of ``cliqueness'' (see \cite{Ertem2016}). \\
Other definitions of the clustering coefficient in multilevel structures have been proposed (see, in particular, \cite{DeDomenico2013,Cozzo2015,Bartesaghi2022}) and all have revealed that an unambiguous extension is not possible in a multidimensional perspective. 
These contributions focus only on undirected networks, hence neglecting directed patterns that could be relevant empirically. \\
Our contribution to this literature is to formulate new local clustering coefficients for weighted directed multiple networks. Since a node in a multilayer network can be clustered under different perspectives, we measure how deep a node is involved in a cohesive structure focusing on a single node, on a single layer or on the entire system. These alternative coefficients convey several characteristics inherent to the complex topology of the multilayer network.\\  
We extend to the multilayer architecture the clustering coefficients already proposed in the literature for weighted directed monoplex networks (see \cite{Fagiolo2007} and \cite{CleGra}). 
Additionally, we generalise to the directed case the coefficient provided in \cite{DeDomenico2013}.

The proposed coefficients have been tested considering international trade data based on the World Input-Output Database (\cite{Timmer}).
Traditionally, these data have been studied considering a monoplex structure (see, e.g., \cite{FagReyes, FagioloSquartini}, \cite{Maluck}, \cite{Piccardi2012, PiccardiPlos}) or a bipartite network (see \cite{Cristelli}, \cite{CingolaniPanz}, \cite{Saracco}). 
However, monoplex networks, based on countries or sectors, and bipartite networks, linking countries to sectors, cannot capture the full spectrum of interactions that commonly arise between countries in the international commodity network. 
Relations between countries in different sectors are also neglected in multiplex trade networks without non-diagonal couplings and where nodes are countries and layers are sectors (see \cite{Mastrandrea}, \cite{Menichetti}, \added{\cite{Alves2018, Alves2019}}).

Trade data integrate different aspects, such as the occurrence of various  sectors, of imports and exports. They can therefore be effectively described by a directed and weighted complex structure, where each layer represents import and export relationships between countries for a given sector. International trade data have been modelled by \cite{Barigozzi2010multinetwork} using a weighted directed network structured on more than one layer, where each layer refers to a commodity-specific network of import/export between countries. Classical topological indicators for directed network have been studied in order to capture the statistical similarities between countries and to track how these commodity-specific networks evolve in time.
\added{Recently, an interesting contribution (\cite{Alves2022}) uses a suitable adaptation of eigenvector centrality in the multilayer framework to measure the economic dominance of buyers and sellers over time.} The study performed by \cite{Ren2020} on the multilayer world trade network in a given time interval reveals the particular nested structure that characterises this set of networks. This property allows to quantify the products’ complexity.\\
To fully take into account the architecture of trade relationships, a multilayer network \deleted{with non-diagonal couplings} is here considered in order to disentangle the effects of countries and sectors and to jointly consider the interactions between them.  Indeed, modelling a complex system as a multilayer network allows to gain information that cannot be captured taking individual layers separately. Results show that the coefficients highlight the role of countries and sectors in the whole trade taking into account both inter-layer and intra-layer flows. In particular, we identify countries that are densely interconnected in all sectors and, at the same time, we bring out countries that are prominent in specific sectors. Additionally, the multilayer structure allows to unveil relevant sectors in terms of interconnection in the whole trade.

The paper is structured as follows. Section \ref{Preliminaries} reports the technical definitions related to multilayer networks. In Section \ref{Triangles in directed multilayer networks} we formally define triangles for multilayer directed networks. Section \ref{Clustering for directed multilayer networks} provides a new formulation of local clustering coefficients by means of supradjacency matrices. In addition, we illustrate how the new coefficients generalise the classical definitions provided in the literature for monoplex networks.  Section \ref{ce} analyses the world trade network in light of the new coefficients. Conclusions follow in Section \ref{Conclusions}.

\section{Preliminaries}
\label{Preliminaries}
In this section we recall some notions and definitions about directed graph \citep{Harary,Bang-Jensen2008} which are essential to introduce the  notation we will use in the paper. 

A directed network $G=(V,E)$ consists of a set $V$ of $N$ vertices (or nodes) and  a set $E\subseteq V\times V$ of ordered pairs of elements of $V$ (arcs or directed edges).  
If there is an arc $(i,j)$ from node $i$ to node $j$, then $i$ and $j$ are said to be adjacent.  
There is a bilateral arc between node $i$ and node $j$ if both $(i,j)\in E,$ and $(j,i)\in E$. 
A directed graph is weighted if a weight $w_{ij} > 0$ is associated with an arc $(i,j)$ in $G$. 
 
We represent the binary adjacency relations between couples of nodes by a square (not necessarily symmetric) matrix $\textbf{A}$ of order $N$ (binary adjacency matrix), whose elements are $a_{ij}=1$ if  $(i,j) \in E$, $0$ otherwise.
The entry $a_{ij} \in \mathbf{A}$ represents the arc outgoing from $i$ and incoming to $j$. Then, the $1$'s on the row $i$ are the arcs that leave the node $i$. 
A weighted directed network is completely described by the real square matrix $\textbf{W}$ of order $N$ (weighted adjacency matrix), whose elements $w_{ij}$ are different from zero if $(i,j)\in E$, and $w_{ij}= 0$ otherwise.

In this paper we deal with node-aligned multiplex networks, with non-diagonal couplings (see \cite{Kivela2014} for the taxonomy of the multidimensional networks).
The assumption that the network is non-diagonal coupled means that there may exist inter-layer arcs not only between nodes and their counterparts, but also between a node $i$ in a given layer and a node $j \neq i$ in a different layer.
\added{For short, we will adopt throughout the paper the terminology multilayer network to denote this kind of network.} A multilayer
network is a family of networks $G_\alpha=(V_\alpha,E_\alpha),\alpha=1,...,L$, where each network $G_\alpha$ represents the layer $\alpha$ and a node $i \in V_\alpha$ is adjacent to $j \in V_\beta$, $\forall \alpha,\beta=1,...,L$ if they are connected by an arc. Since the multilevel network is node-aligned, nodes are repeated across all levels, i.e. $V_\alpha=V_\beta=V,\ \forall \alpha,\beta=1,...,L$. Furthermore $E_\alpha$ identifies all the arcs connecting nodes in the layer $\alpha$ to nodes in the same layer (intra-layer connections) and to nodes in all the other layers (inter-layer connections). In particular, we will denote with $E_{\alpha \beta}$ the set of arcs from layer $\alpha$ to $\beta$, i.e. $E_\alpha=\bigcup_\beta E_{\alpha \beta}$.

\deleted{In what follows, we call this kind of networks multiplex networks or simply multiplex.}

\noindent A weight $w^{\left[\alpha\beta\right]}_{ij} > 0$ is associated with an arc $(i,j)$ in $E_{\alpha \beta}$, with $\alpha, \beta = 1,\dots,L$. Note that when $\alpha=\beta$ we intend that there is a weighted arc $w^{\left[\alpha \alpha\right]}_{ij} > 0$ between nodes $i$ and $j$ in the layer $\alpha$. 

A convenient way to represent the adjacency relations between pair of nodes is represented by the supradjacency matrix. It is defined as a matrix, with $L\times L$ square blocks of order $N$:

\begin{equation}
\label{supra}
	\mathbf{W}=
	\begin{bmatrix}
		\mathbf{W}^{\left[11\right]} & \mathbf{W}^{\left[12\right]} & \cdots & \mathbf{W}^{\left[1L\right]} \\
		\mathbf{W}^{\left[21\right]} & \mathbf{W}^{\left[22\right]} & \cdots & \mathbf{W}^{\left[2L\right]} \\
		\vdots & \vdots & \ddots & \vdots \\
		\mathbf{W}^{\left[L1\right]} & \mathbf{W}^{\left[L2\right]} & \cdots & \mathbf{W}^{\left[LL\right]}
	\end{bmatrix}
\end{equation}

\noindent where $\textbf{W}^{\left[\alpha \alpha \right]}$, $\alpha=1,...,L$, are the weighted adjacency relations in the layer $\alpha$, whereas $\textbf{W}^{\left[\alpha \beta\right]}$ are the weighted adjacency relations between nodes on layers $\alpha$ and nodes on layer $\beta$. We denote the unweighted version by $\mathbf{A}$.

\noindent We denote the generic element of $	\mathbf{W}$ as $w_{hk}$ with $h,k=1,...,NL$, where 
\begin{equation}
	\label{index}
	h=N(\alpha-1)+i, k=N(\beta-1)+j.
\end{equation} 
Notice that the indices $h,k$ identify the position in the supradjacency matrix $\mathbf{W}$ of the weight of the arc $(i,j) \in E_{\alpha,\beta}$ (i.e. $w^{\left[\alpha\beta\right]}_{ij}=w_{hk}$). Thus, from now on, we always assume this relation between $h,k$ and $\alpha,\beta,i,j$. 

\noindent The in-degree $d^{[\alpha]}_{i,in}$ of a node $i$ on layer $\alpha$ is the number of arcs pointing towards $i$ from any layer. This degree can be expressed as follows:
\begin{equation}\label{in_deg}
	d^{[\alpha]}_{i,in}=(\textbf{A}^T\textbf{1})_h
\end{equation}
where $\textbf{1}$ is the $NL$-vector of ones. 
Definition \eqref{in_deg} represents the $h$-th component of the in-degree vector, with $h$ as in \eqref{index}. By the definition of the matrix $	\mathbf{A}$, $d^{[\alpha]}_{i,in}$  is obtained as the sum of the in-degree related to arcs in the same layer $\alpha$ and the in-degrees related to incoming arcs from other layers $\beta$.

\noindent Similarly, the out-degree is:
\begin{equation}\label{out_deg}
	d^{[\alpha]}_{i,out}=(\textbf{A}\textbf{1})_h.
\end{equation}


\noindent The degree  $d^{[\alpha]}_i$ of a vertex $i$ in the layer $\alpha$ is then:
\begin{equation}\label{tot_deg}
	d^{[\alpha]}_i=d^{[\alpha]}_{i,in}+d^{[\alpha]}_{i,out}=[(\textbf{A}^T+\textbf{A})\textbf{1}]_h.
\end{equation}

\noindent Bilateral arcs between the node $i$ and its adjacent nodes, if any, are represented as:
\begin{equation}\label{bil_deg}
	d^{\left[\alpha\right]}_{i,\leftrightarrow}=\left(\textbf{A}^{2}\right)_{hh}.
\end{equation}

\noindent The in-degree of a node $i$ with respect to all layers is defined as:
\begin{equation}\label{totin_deg}
		d_{i,in}=\sum_{\alpha=1}^{L}d^{[\alpha]}_{i,in}.
\end{equation}
Out-degree $d_{i,out}$ 
with respect to all layers is defined similarly. 
The total degree on the multilayer is then $d_{i}=d_{i,in}+d_{i,out}$. 


Moving to the weighted case, the previous definitions can be replaced by the strength of a node $i$:
\begin{equation}\label{in_str}
	s^{[\alpha]}_{i,in}=(\mathbf{W}^T\mathbf{1})_h
\end{equation}

\begin{equation}\label{in_out}
	s^{[\alpha]}_{i,out}=(\textbf{W}\textbf{1})_h
\end{equation}

\noindent The strength  $s^{[\alpha]}_i$ of a vertex $i$ in the layer $\alpha$ is then:
\begin{equation}\label{str}
	s^{[\alpha]}_i=s^{[\alpha]}_{i,in}+s^{[\alpha]}_{i,out}=[(\textbf{W}^T+\textbf{W})\textbf{1}]_h.
\end{equation}

\noindent The in-strength of a node $i$ with respect to all layers is defined as:
\begin{equation}\label{totin_str}
	s_{i,in}=\sum_{\alpha=1}^{L}s^{[\alpha]}_{i,in}.
\end{equation}
Out-strength $s_{i,out}$ with respect to all layers is defined similarly. The total strength of $i$ is then $s_{i}=s_{i,in}+s_{i,out}$. 

We define the strength related to bilateral arcs between the node $i$ and its adjacent nodes as:

\begin{equation}\label{bil_str}
	s^{\left[\alpha\right]}_{i,\leftrightarrow}=\frac{(\textbf{WA}+\textbf{AW})_{hh}}{2}.
\end{equation}

\noindent 
Formula (\ref{bil_str}) extends formula (\ref{bil_deg}) to the weighted case, multiplying each bilateral link by the arithmetic mean of its weights\footnote{Other choices are possible but could be not equally effectiveness in computation.}.
With this choice, we are assuming that the bilateral strength sums the average weight of each bilateral arc.

\section{Triangles in directed multilayer networks}
\label{Triangles in directed multilayer networks}

To formally provide a definition of clustering coefficients in multilayer context, it is worth at first to introduce the definition of triangle.

In a multilayer, we define a triangle as a triad of nodes $i,j,k$ laying up to three different layers. Vertices in triangles can then be linked by both inter and intra-layer arcs, irrespective of their orientation. In this way, we can take into account all possible closed triads, in all inter or intra-layer directions\footnote{In directed multilayer networks, triangles around a given node $i$ on layer $\alpha$ can be classified in \textit{out, in, cycle and middleman} triangles.}. This definition extends to the directed case the one adopted in \cite{Bartesaghi2022}. \deleted{for multiplex weighted networks}

The number of triangles 
can be computed by counting the oriented $3$-cycles around the focal node. Similarly, in a weighted directed multilayer network a weight can be given to a triangle 
for instance by simply multiplying the weights of its three arcs. 
This is not a univocal choice, and other ways to attribute weights have been proposed in the literature, giving rise to different clustering coefficients in this context.


The total number $T_i^{[\alpha]}$ 
of real triangles for the node $i$ on a layer $\alpha$ is provided by the following expression in terms of supradjacency matrix $\mathbf{A}$: 

\begin{equation}\label{Triangles}
T_i^{[\alpha]} = \frac{1}{2}(\left( \mathbf{A}+\mathbf{A}^T\right)^3)_{hh}
\end{equation}

Formula \eqref{Triangles} allows to extend the definitions on each layer and on the whole network. 
Indeed, the number of triangles for the node $i$ on any layer is then $T_i=\sum_{\alpha=1}^{L}T_i^{[\alpha]}$, and the total number of triangles in a single layer $\alpha$ is $T^{[\alpha]}=\sum_{i=1}^{N}T_i^{[\alpha]}$, finally the total  number of triangles in the network is $T=\sum_{i=1}^{N}\sum_{\alpha=1}^{L}T_i^{[\alpha]}$.

\begin{figure}[H]
	\centering
	\includegraphics[scale=0.4]{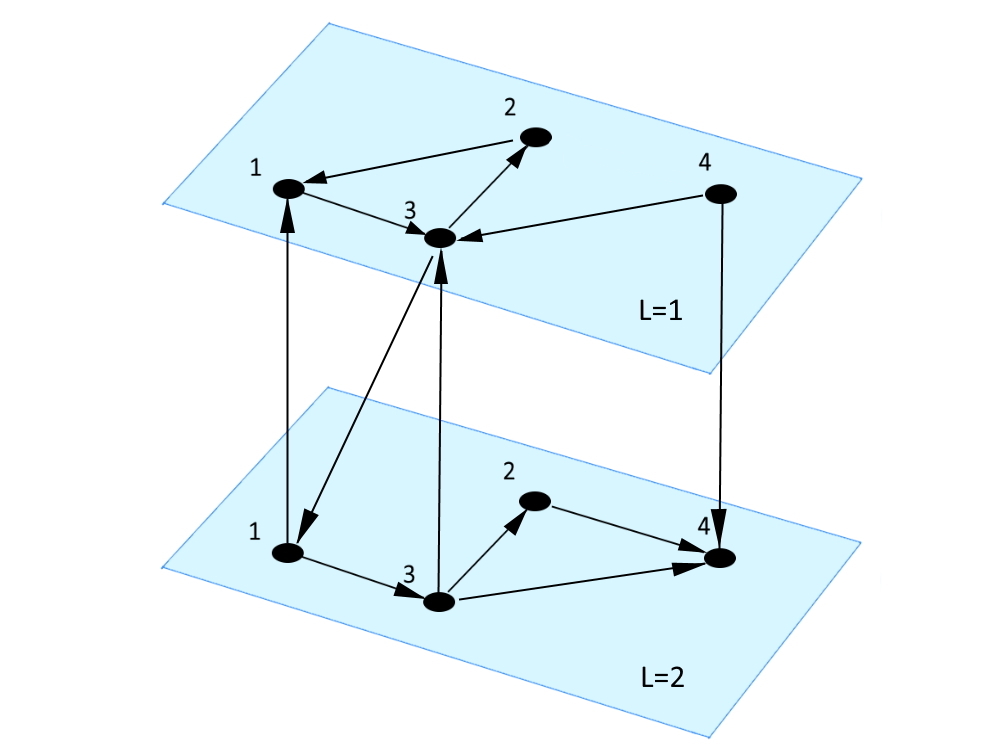}
	\caption{Example of a binary directed multilayer network with 4 nodes and 2 layers}
	\label{fig2}
\end{figure}

Figure \ref{fig2} illustrates what has been introduced so far on a simple example. Four nodes are connected by oriented arcs on two distinct layers. For the sake of simplicity, there are no bilateral arcs and all the weights are set equal to $1$.
 
For instance we can observe that on layer $1$ node $3$ belongs to three directed triangles, whereas on layer $2$ node $3$ belongs to only \added{two triangles}. Similarly, we can calculate for each node on each level the number of actual and potential triangles.

\section{Clustering coefficients in weighted directed multilayer networks}
\label{Clustering for directed multilayer networks}

We provide here the mathematical definitions of clustering coefficients for multilayer directed networks.
In this context, we can define four different types of clustering coefficients: three coefficients, depending on the node and/or layer under consideration, and a global one, relating to the entire network. 
Moreover, in a weighted and directed multilayer network, triangles can be weighted in more than one way, giving rise to more than one possible definition of weighted clustering coefficient.

A first possibility is to consider all actual directed triangles that a node $i$ forms with its neighbours, where the triangle's weight is the average of the weights of the links  linking a node $i$ to its adjacent nodes $j$ and $k$. This is in line with the coefficient introduced in \cite{CleGra} for monoplex networks.\\ 
Hence, we define the coefficient $C(i,\alpha)$ for a node $i$ in the level $\alpha$ as:

\begin{equation}
\label{CleGra}
C(i,\alpha)=\frac{\big[ \left( \mathbf{W}+\mathbf{W}^{T} \right)\left(\mathbf{A}+\mathbf{A}^{T} \right)^{2}\big]_{hh}}{2\left[ s^{[\alpha]}_i\big(d^{[\alpha]}_i-1\big)-2s^{\left[\alpha\right]}_{i,\leftrightarrow}\right] }.
\end{equation}

\noindent where $\mathbf{W}$ is defined as in \eqref{supra} and $\mathbf{A}$ accordingly, $d^{[\alpha]}_i,s^{[\alpha]}_i,s^{\left[\alpha\right]}_{i,\leftrightarrow}$ are defined as in formulas \eqref{tot_deg}, \eqref{str} and \eqref{bil_str} and $h$ is the position index defined in \eqref{index}. \\
In particular, the numerator extends to the weighted case formula \eqref{Triangles} that counts the number of directed triangles $T_i^{[\alpha]}$. The denominator provides all possible (weighted) directed triangles that $i$ could form in or outside the layer $\alpha$. 

An alternative way to weight the triangles is to use the geometric mean of the three weights of the arcs. This choice is in agreement with the coefficient proposed by \cite{Fagiolo2007} for monoplex networks. Hence, we define the coefficient, denoted by $\hat{C}(i,\alpha)$, as:

\begin{equation}
\label{Fagiolo}
\hat{C}(i,\alpha)=\frac{\left(\mathbf{\hat{W}}+\mathbf{\hat{W}}^{T}\right)^3_{hh}}{2\left[ d^{[\alpha]}_i\big(d^{[\alpha]}_i-1\big)-2d^{\left[\alpha\right]}_{i,\leftrightarrow}\right]}.
\end{equation}

\noindent where $\mathbf{\hat{W}}$ is the matrix whose entries 
are $\hat{w}^{[\alpha\beta]}_{ij}=\left(\frac{w^{[\alpha\beta]}_{ij}}{max(w^{[\alpha\beta]}_{ij})}\right)^{\frac{1}{3}} \ \forall i,j,\alpha,\beta$, being ${w}^{[\alpha\beta]}_{ij}$ defined as in Section \ref{Preliminaries}.
Elements  $d^{[\alpha]}_i,d^{\left[\alpha\right]}_{i,\leftrightarrow}$ are defined as in formulas \eqref{tot_deg} and \eqref{bil_deg} and $h$ is the position index defined as in \eqref{index}.

It is noticeable that, in \eqref{CleGra} and \eqref{Fagiolo}, bilateral arcs $2s^{\left[\alpha\right]}_{i,\leftrightarrow}$ and $2d^{\left[\alpha\right]}_{i,\leftrightarrow}$ are removed being ``false'' triangles, formed by $i$ and by a pair of directed arcs pointing to the same node, e.g., $i \rightarrow j$ and $j \rightarrow i$. Notice that in a directed framework a node can form up to two triangles with each pair, also including two ``false'' potential triangles for each bilateral link.

A third possibility to assign a weight to actual and potential triangles can be found in the recent literature (see \cite{DeDomenico2013} and \cite{Jia2021}). It consists in taking the product of the three weights of the symmetrised arcs, assuming that the adjacency matrix is normalised as for the coefficient in formula (\ref{Fagiolo}). According to this choice, the clustering coefficient can be defined as

\begin{equation}
\label{DeDom}
\tilde{C}(i,\alpha)=\frac{\left(\mathbf{\tilde{W}}+\mathbf{\tilde{W}}^{T}\right)^3_{hh}}{ \left[s^{[\alpha]}_{i}\right]^{2}-\sum_{k\neq h}(w_{hk}+w_{kh})^2}
\end{equation}

\noindent where $\mathbf{\tilde{W}}$ is the normalised matrix whose entries are $\tilde{w}^{[\alpha\beta]}_{ij}=\frac{w^{[\alpha\beta]}_{ij}}{max(w^{[\alpha\beta]}_{ij})} \ \forall i,j,\alpha,\beta$, and $h, k$ are the position indices defined as in \eqref{index}. The denominator in formula (\ref{DeDom}) accounts for the weights of all the connected triads centred in the node $i$ on level $\alpha$ and it can be considered the weight of all the potential triangles. Let us notice that to the (missing or actual) arc opposite to the node of interest is assigned a weight equal to $1$. Of course, $s^{[\alpha]}_i$ is again the total strength of the node $i$ in level $\alpha$ defined as in formula \eqref{str}.

Formulas (\ref{CleGra}), (\ref{Fagiolo}) and (\ref{DeDom}) provide local coefficients, since they refer to a single node in a single layer. It turns out to be useful to define coefficients that account for an intermediate clustering structure of the network, both considering homologous nodes on different layers or all the nodes within a given layer. It is also possible to introduce an average coefficient on the whole network that traces the concept of transitivity, well-known for monolayer networks.

\noindent In order to introduce these further generalizations, we refer to formula (\ref{CleGra}) but quite naturally similar extensions of coefficients (\ref{Fagiolo}) and (\ref{DeDom}) can be provided.

First we can consider a single node on all the layers on which it lies. This involves calculating the total weight of triangles to which these homologous nodes belong and dividing by the potential ones:

\begin{equation}
\label{CleGra_node}
C_{N}(i)=\frac{\sum_{\alpha =1}^{L}\big[ \left( \mathbf{W}+\mathbf{W}^{T} \right)\left(\mathbf{A}+\mathbf{A}^{T} \right)^{2}\big]_{hh}}{2\sum_{\alpha =1}^{L}\left[ s^{[\alpha]}_i\big(d^{[\alpha]}_i-1\big)-2s^{\left[\alpha\right]}_{i,\leftrightarrow}\right]}.
\end{equation}

\noindent where $h=N(\alpha-1)+i$ (by formula (\ref{index})) and we sum triangles over $\alpha=1, \dots, L$.\\
It is noteworthy that $C_{N}(i)$ can be obtained as the weighted mean of the local coefficients $C(i,\alpha)$ defined in formula (\ref{CleGra}). Each local coefficient has a weight in the weighted average equal to 
$$\frac{ s^{[\alpha]}_i\big(d^{[\alpha]}_i-1\big)-2s^{\left[\alpha\right]}_{i,\leftrightarrow}}{\sum_{\alpha =1}^{L}\left[ s^{[\alpha]}_i\big(d^{[\alpha]}_i-1\big)-2s^{\left[\alpha\right]}_{i,\leftrightarrow}\right]}.$$ 
In other words, the more the strength (and the degree) of the node $i$ in the layer has a high incidence on the total strength (and degree) of the node in all the layers, the more $C(i,\alpha)$ affects $C_{N}(i)$.

Alternatively, we can focus on a single layer and define a coefficient regarding all the nodes in that layer. In particular, these nodes contribute to the coefficient on the basis of all their arcs, within the layer or outside:

\begin{equation}
\label{CleGra_level}
C_{L}(\alpha)=\frac{\sum_{i =1}^{N}\big[ \left( \mathbf{W}+\mathbf{W}^{T} \right)\left(\mathbf{A}+\mathbf{A}^{T} \right)^{2}\big]_{hh}}{2\sum_{i =1}^{N}\left[ s^{[\alpha]}_i\big(d^{[\alpha]}_i-1\big)-2s^{\left[\alpha\right]}_{i,\leftrightarrow}\right] }.
\end{equation}

\noindent where, again, $h=N(\alpha-1)+i$ and the sums are over $i=1, \dots, N$. \\
\noindent Also in this case, the clustering coefficient for a layer can be written as the weighted mean of the local coefficients $C(i,\alpha)$. The incidence of $C(i,\alpha)$ on $C_{L}(\alpha)$ depends on the relation between the strength and the degree of the node $i$ in the layer $\alpha$ and the total strength and degree of all nodes in the layer.


Finally, we can take into account all nodes on all levels, with all their actual or potential triangles:

\begin{equation}
\label{CleGra_network}
C=\frac{\sum_{h =1}^{NL}\big[ \left( \mathbf{W}+\mathbf{W}^{T} \right)\left(\mathbf{A}+\mathbf{A}^{T} \right)^{2}\big]_{hh}}{2\sum_{i =1}^{N}\sum_{\alpha =1}^{L}\left[ s^{[\alpha]}_i\big(d^{[\alpha]}_i-1\big)-2s^{\left[\alpha\right]}_{i,\leftrightarrow}\right]}.
\end{equation}

Note that formula (\ref{CleGra_network}) extends the definition of transitivity for the entire network to the directed multilayer case.

\section{Computational experiment}\label{ce}

\subsection{Dataset, multilayer network and preliminary analyses}
\label{sec:data}
To test the effectiveness of the proposed coefficients, we run a computational experiment applying them to real-world data. 
We refer to the international trade data collected in the World Input-Output Database (WIOD) 2016 Release (\cite{Timmer}). WIOD consists of a series of databases and covers 28 EU countries and 15 other major countries. Including the biggest countries
in the world, this set covers more than 85 per cent of world GDP (see, e.g., \cite{Timmer}).  However, to complete the data and make them suitable for different modelling purposes, it is also considered a region called the Rest of the World (RoW) that collects all other countries in the world. In Table \ref{table_WIOD_countries}, we display the list of countries that are included in our analysis.
The WIOD contains annual time-series of world input–output tables covering the period from 2000 to 2014. Data consider bilateral trades of exports, expressed in millions of U.S. dollars, for each couple of origin and destination countries and distinct between 56 sectors (see Table \ref{table_WIOD_industries} for a list of sectors). Sectors are classified according to the International Standard Industrial Classification (revision 4) (see \cite{UN}). In this analysis, we focus on the data of last available year, namely, 2014. \\
By these data we construct a multilayer weighted directed network, where each layer is represented by a sector. In each layer, a node corresponds to a country and a  weighted arc represents the total amount of incoming or out-coming flows between a couple of countries in the same sector. In the same layer, we exclude self-loops, i.e. transaction of a country in the same sector. Inter-layer arcs are instead weighted with the amount of imports or exports between (the same or different) countries in different sectors. We obtain a multilayer network with 44 nodes and 56 sectors, represented by an asymmetric and weighted supradjacency matrix with 2464 rows and columns. \\

\begin{figure}[!h]
	\begin{center}
		\includegraphics[height=8cm,width=14cm]{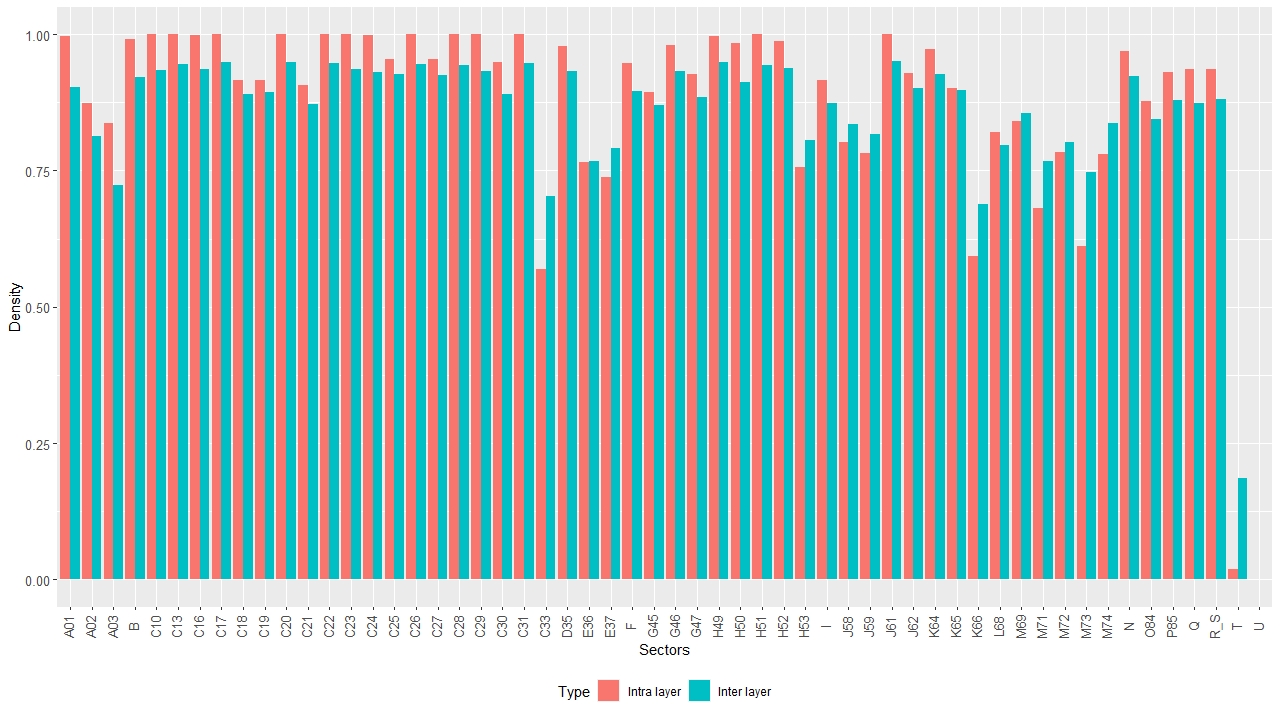}
		\caption{Intra-layer and inter-layer density of the multilayer network for each sector}
		\label{fig:Dens}
	\end{center}
\end{figure}

To give a first idea of the network structure, we display in Figure 	\ref{fig:Dens} the density of each layer and the average inter-layer density. The average inter-layer density of a sector has been computed as the average value of the densities of the graphs characterised by the connections between the sector and each other layer. It is noticeable how several sectors show a high internal density in the layer, while a lower inter-layer density is observed on average. Sector ``T" \added{\emph{(Activities of households)}} ) is instead very sparse and tends to be more connected with other layers. Sector ``U" \added{\emph{(Activities of extraterritorial organizations and bodies)}} is completely isolated having neither internal connections nor connections with other layers. Therefore, we decided to remove this layer from the network and to deal with 55 sectors.

\begin{figure}[h!]
	\begin{center}
		\includegraphics[height=12cm,width=14cm]{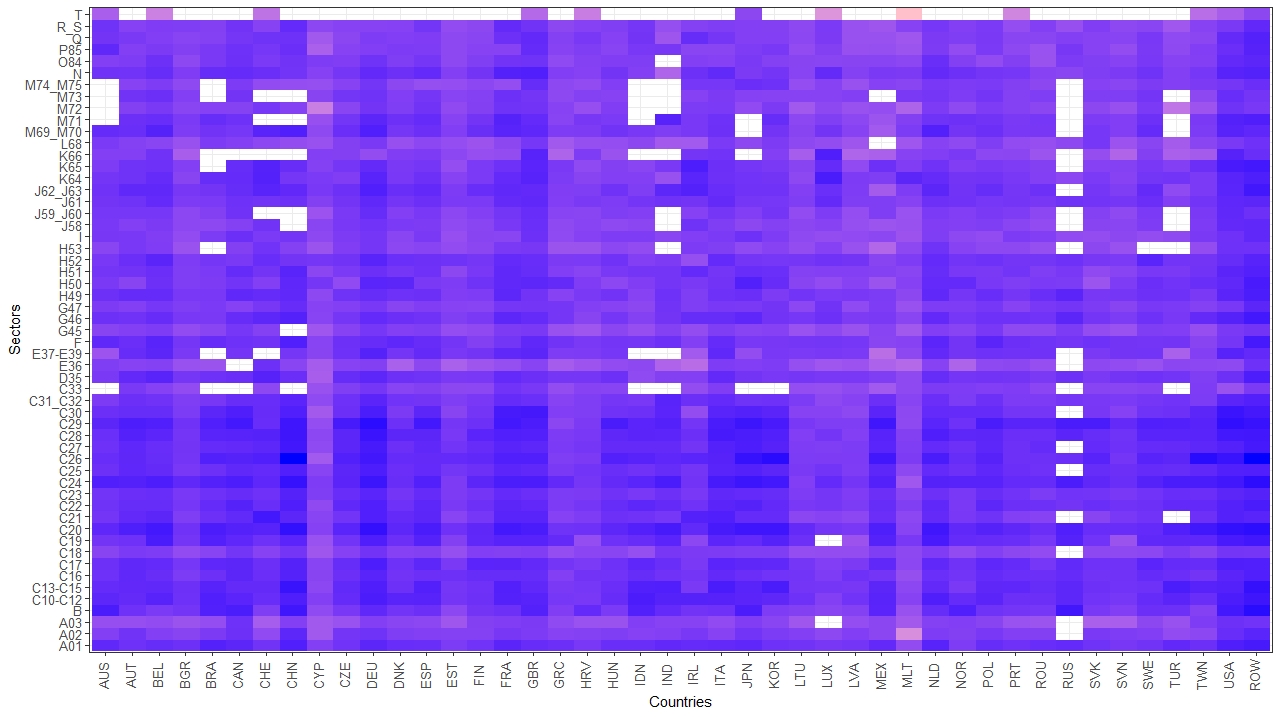}
		\caption{Total strength for each sector (layer) and country (node). The plot considers only intra-layer flows. Darker colours are associated with higher values of strength. Empty dots are related to combination of sectors and countries where no arcs are observed.}
		\label{fig:St}
	\end{center}
\end{figure}

We report in Figure \ref{fig:St} the total strength for each combination of sector and country. The total strength has been obtained as the sum of in and out strength of the country in the layer. It is interesting to note how manufacturing sectors (from ``C10" to ``C33" in the plot) are characterised by the highest volumes. In terms of country, China, USA and ROW show the highest strengths. On the one hand, China tends to be the dominant country in several sectors but shows also a strength equal to zero in specific sectors (as \added{Repair and installation of machinery and equipment} ``C33", \added{Wholesale and retail trade and repair of motor vehicles and motorcycles} ``G45", \added{Publishing activities} ``J58", etc.). On the other hand, USA and ROW have  average strength a bit lower than China but these countries show connections in all sectors.

\begin{figure}[H]
	\begin{center}
		\includegraphics[height=7cm,width=12cm]{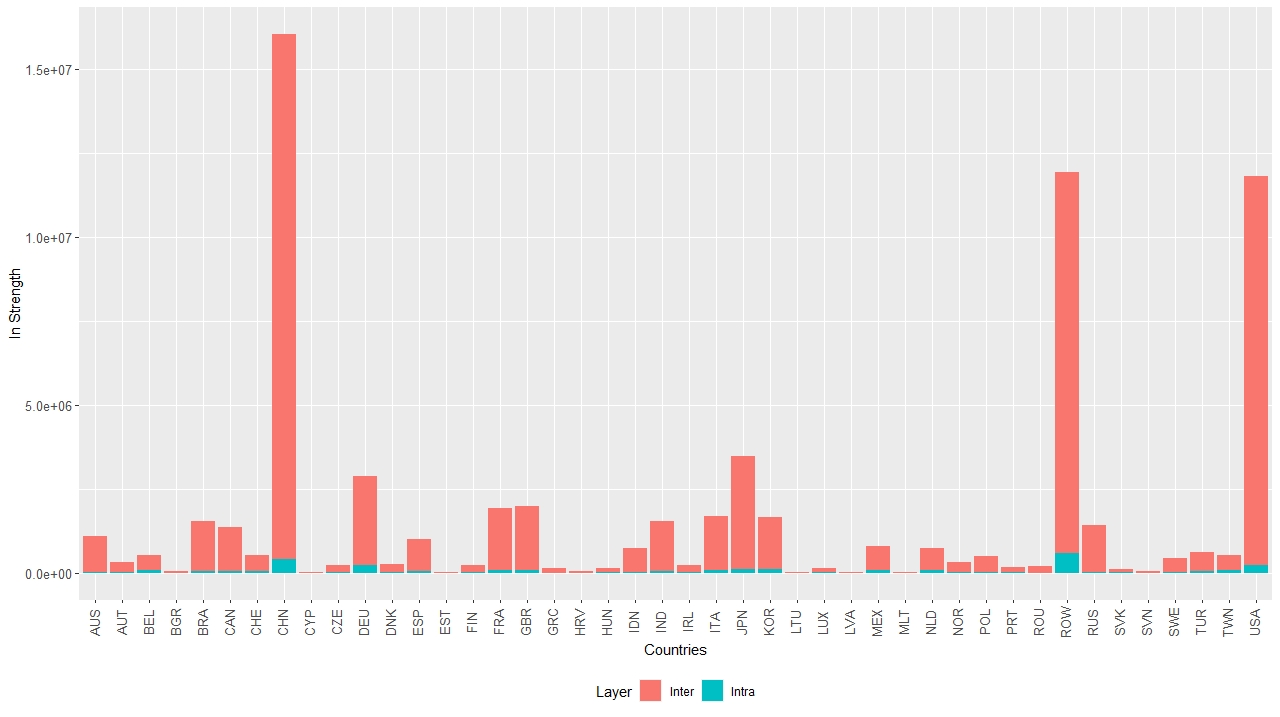}
		\caption{\added{\lq\lq In\rq\rq strength for each country differentiated between intra-layer and inter-layer connections.}}
		\label{fig:StDir}
	\end{center}
\end{figure}

\begin{figure}[H]
	\begin{center}
		\includegraphics[height=7cm,width=12cm]{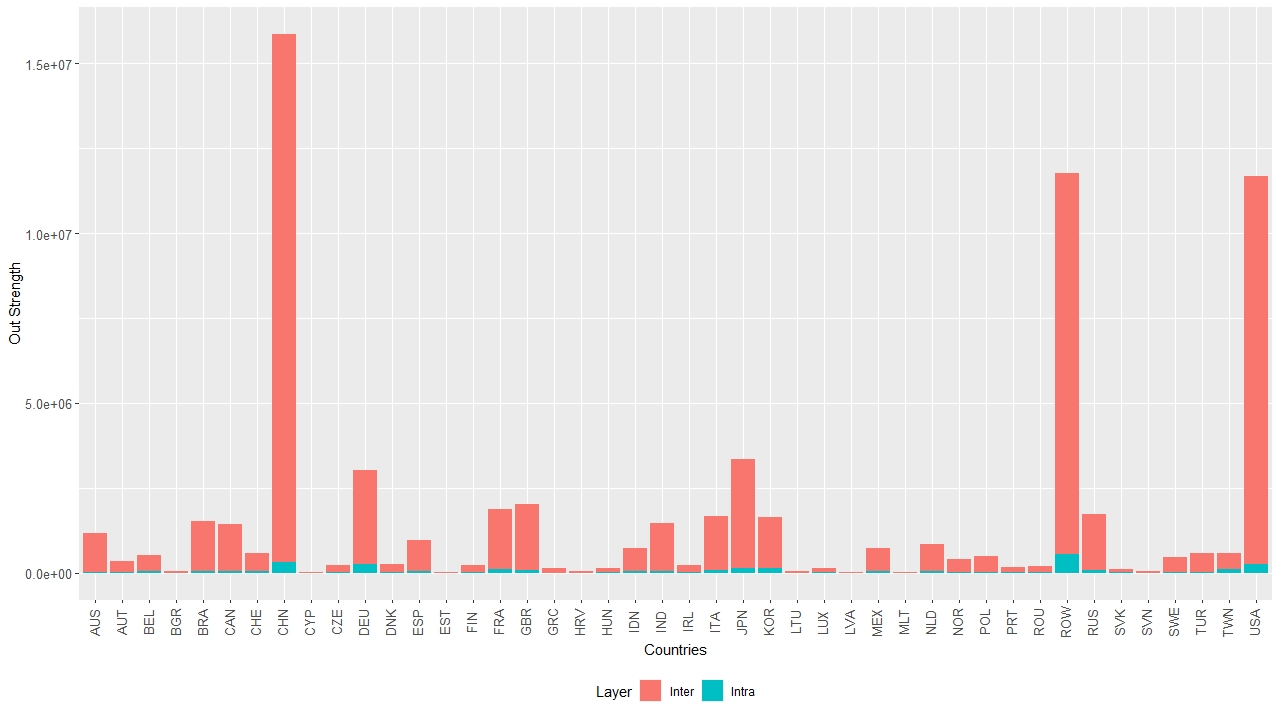}
		\caption{\added{\lq\lq Out\rq\rq strength for each country differentiated between intra-layer and inter-layer connections.}}
		\label{fig:StDirO}
	\end{center}
\end{figure}

A specific focus on countries' strength is made \added{in Figures \ref{fig:StDir} and \ref{fig:StDirO}}. In particular, we distinguish between in and out strength and between strength related to inter-layer and intra-layer connections. A high volume of trade due to inter-layer connections is observed. In particular, USA, AUS and CHN show the highest exposition towards inter-layer arcs. We have values lower than 3\% for the ratio between the strength related to arcs in the same layer and the strength due to arcs that connect different layers. Smaller countries (as MLT, LUX, TWN, SVK, HUN) are instead more concentrated on intra-layer flows. \\
According to imports and exports, NOR, RUS, NLD are characterised by a positive trade balance with a ratio between in and out strength around 80\%. Vice versa MLT, HUN and MEX have the opposite behaviour with the highest ratios (more than 110\%).


\begin{figure}[H]
	\begin{center}
		\includegraphics[height=7cm,width=15cm]{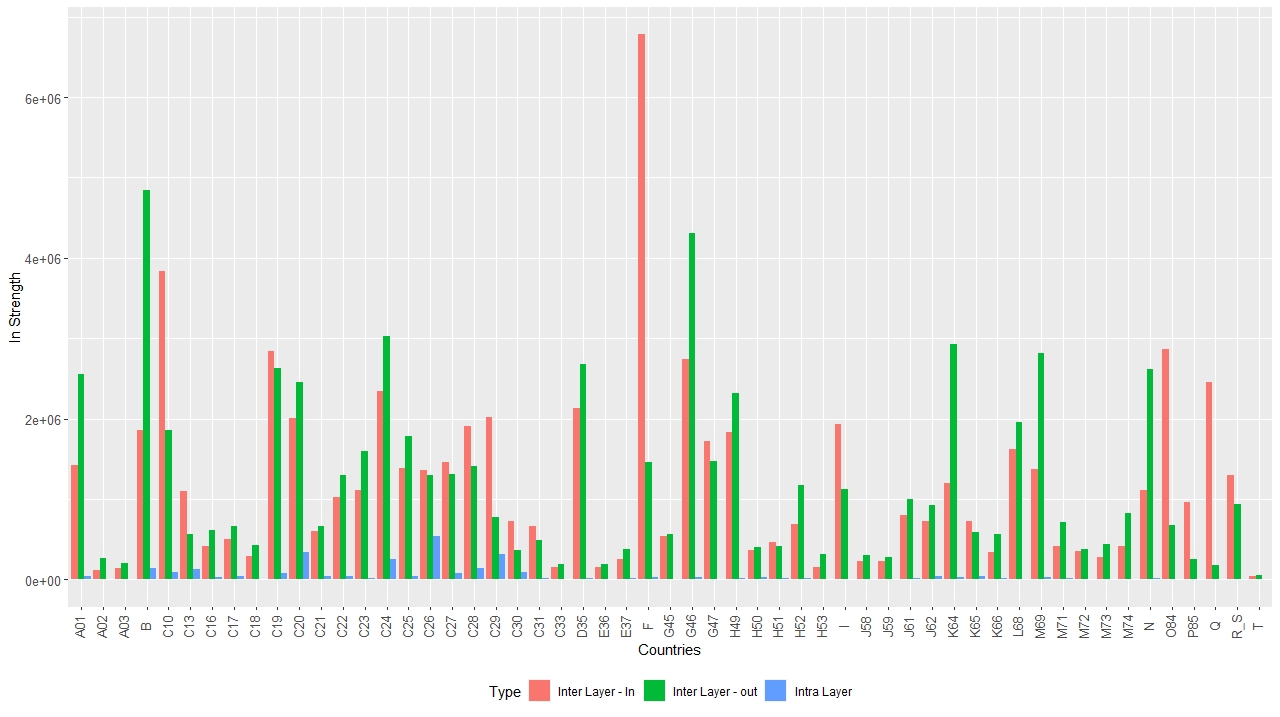}
		\caption{Intra-layer strength for each sector (i.e. transactions in the same sectors). Inter-layer \lq\lq in\rq\rq and \lq\lq out\rq\rq strength for each sector; in this case, we consider directed flows between different sectors.}
		\label{fig:StDirSec}
	\end{center}
\end{figure}

In Figure \ref{fig:StDirSec} we focus on sectors' strength. It is noticeable a very high in-strength for the sector ``F" (\emph{Constructions}), due in particular to directed connections with some manufacturing sectors. Important out-flows are instead observed for the sector ``B" (\emph{Mining and quarrying}), probably due to the use of these materials in other sectors. Finally, we observe how the sector ``C26" (\emph{Manufacture of computer, electronic and optical products}) is characterised by the highest intra-layer transactions, because of important trades between countries.

\subsection{Main Results}
\begin{figure}[H]
	\begin{center}
		\includegraphics[height=6cm,width=16cm]{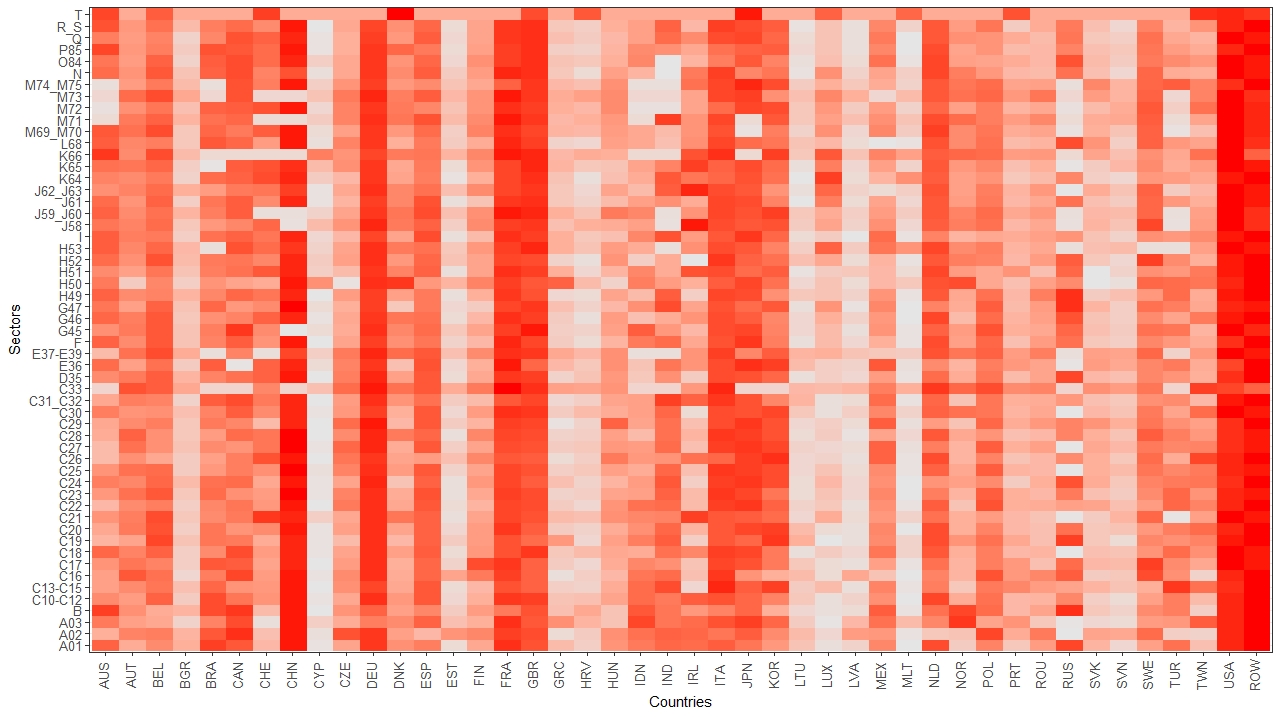}
		\includegraphics[height=6cm,width=16cm]{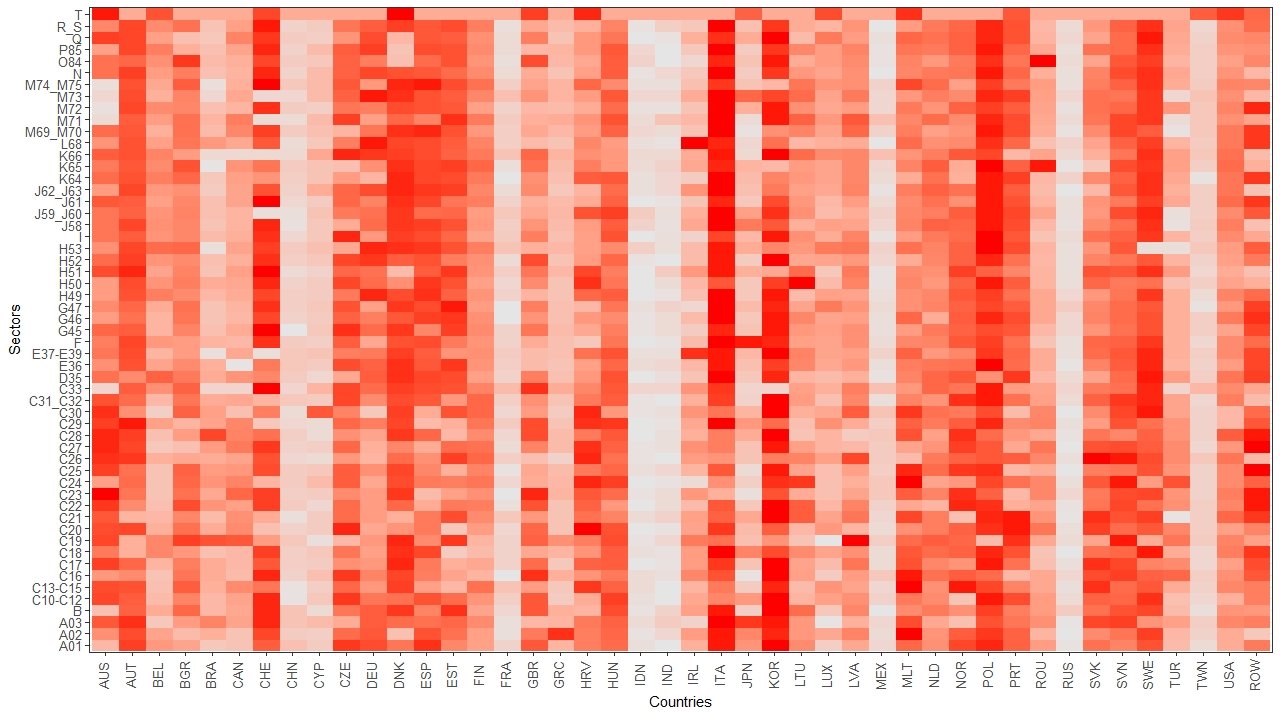}
		\includegraphics[height=6cm,width=16cm]{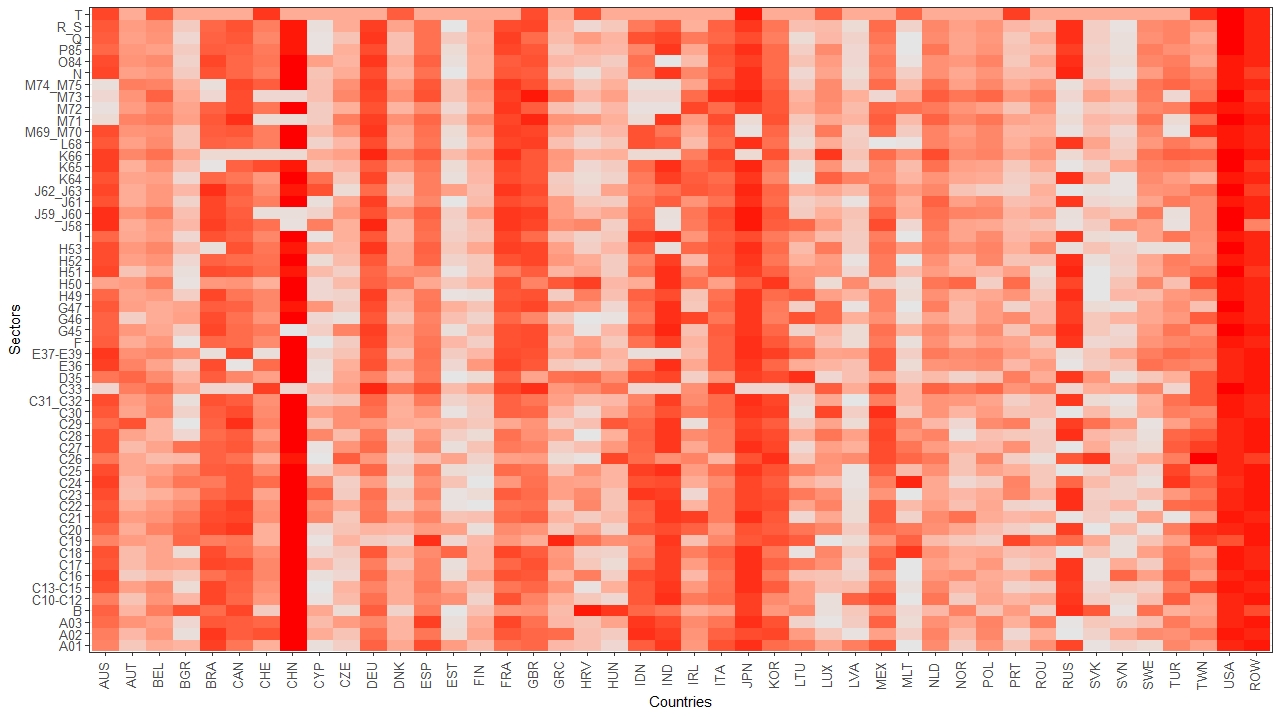}
		\caption{Ranking of clustering coefficients $\hat{C}(i,a)$, $C(i,a)$ and $\tilde{C}(i,a)$ for each sector. Darker red means higher ranking.}
		\label{fig:Clust}
	\end{center}
\end{figure}

We test on the multilayer network the coefficients introduced in Section \ref{Clustering for directed multilayer networks}. We compare at first the local coefficients provided by formulas (\ref{CleGra}), (\ref{Fagiolo}) and (\ref{DeDom}). To this end, we display in Figure \ref{fig:Clust} for each sector the ranking based on the local clustering coefficients. It allows to emphasize which countries appear prominent in terms of interconnections in each sector and, at the same time, we can appreciate different patterns between coefficients. We notice that coefficients $\hat{C}(i,a)$ and $\tilde{C}(i,a)$ tend to provide similar rankings, while coefficients $C(i,a)$ behave in a different way. However, it has to be stressed that values of $C(i,a)$ are very close to one and hence the ranking is in this case not so interesting because very limited differences are observed between countries. Main justification is related to the fact that this coefficient is affected by the number of triangles rather than by their weights. Therefore, given the high density of the network, very few differences can be noticed between countries. Focusing on rankings given by $\hat{C}(i,a)$ and $\tilde{C}(i,a)$, it is noticeable how CHN, USA and ROW appear highly interconnected. In particular, CHN have the highest ranking in the largest number of sectors (27 and 38 sectors according to $\hat{C}(i,a)$ and $\tilde{C}(i,a)$, respectively), but USA and ROW are well connected in all sectors. We have indeed that these two countries belong to the top quartile of the clustering distribution in all sectors. We have instead that CHN shows a very low ranking in specific sectors where it is not well represented. In terms of average ranking, these countries are then followed by FRA, GBR, JAP, DEU. In particular, FRA and GBR tend to be well clustered in several sectors (especially considering the coefficient $\hat{C}(i,a)$), while JAP and DEU are instead not covered in some specific sectors. In particular, FRA has the highest ranking in sector ``C33" (\emph{Air transport}). France is indeed one of the largest and relevant aviation markets in Europe, because of traffic due to its size and geographic location and being home of some of the industry’s flagship names. At a lower average ranking we find ITA, RUS, KOR, NLD, ESP, countries that belong to the top quartile in specific sectors but also show a lower clustering in other sectors. It is also noteworthy how Denmark appears as a top country in sector ``T", \emph{Activities of households}, and Taiwan is highly interconnected in sector ``C26" \emph{Manufacture of computer, electronic and optical products}. Electronic component manufacturing is indeed a pillar of Taiwan's economy, and its role is increasing over time also thanks to the development of technology.

\newpage

\begin{figure}[H]
	\begin{center}
		\includegraphics[height=7cm,width=15cm]{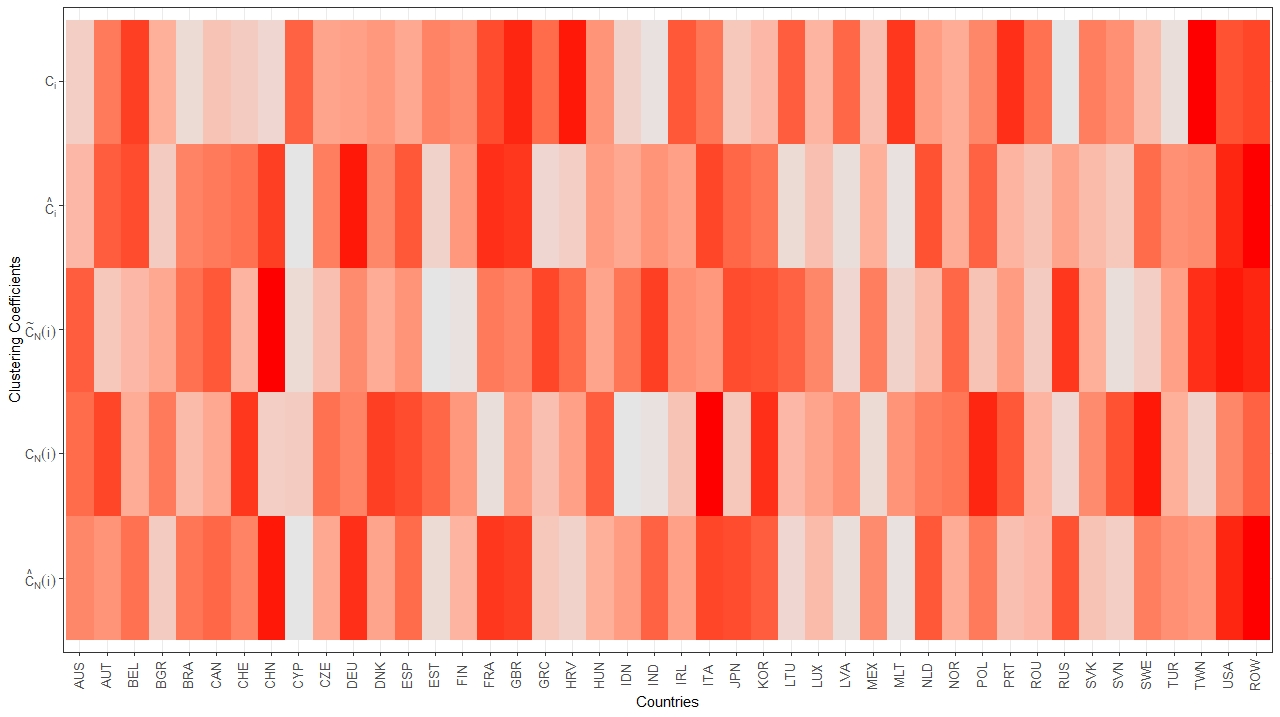}
		\caption{Comparison of rankings of clustering coefficients $\hat{C}_{N}(i)$, $C_{N}(i)$ and $\tilde{C}_{N}(i)$ computed for each country $i$. The comparison is also extended to the average clustering based on monoplex coefficients. In particular $\hat{C}_{i}$ and $C_{i}$ are the average values of the clustering coefficients of each country based on formulas  provided in \cite{Fagiolo2007} and \cite{CleGra} for weighted and directed networks. Darker red means higher ranking.}
		\label{fig:ClustC}
	\end{center}
\end{figure}

To stress the different behaviour of the alternative  clustering coefficients, we show in Figure \ref{fig:ClustC} the countries' ranking computed on the base of the three multilayer coefficients. To get these rankings, we considered the  coefficients of the single node $i$ over all the layers  (defined by formula (\ref{CleGra_node})). We extend the comparison to the local clustering coefficients for monoplex networks. In particular, we compute on each layer a coefficient $\hat{C}_{i}$, based on the formula provided by \cite{Fagiolo2007} and then we average the results between the layers. Similarly, the same procedure has been applied in order to obtain the coefficient $C_{i}$ based on the formula provided in \cite{CleGra}. Therefore, inter-layer connections are not considered in the computation of $\hat{C}_{i}$ and $C_{i}$.  \\
Interestingly, the inclusion of inter-layer effects produces significant disparities in the ranking. This pattern is also justified by the fact that the inter-layer connections are relevant, as shown by the behaviour of density and strength in the previous section. The coefficients referred to monoplex networks show indeed a positive rank correlation but far from one, with the results obtained from multilayer formulas. Finally, we notice a higher level of similarity between $\hat{C}_{N}(i)$ and $\hat{C}_{i}$ than between $C_{N}(i)$ and $C_{i}$.

\begin{figure}[H]
	\begin{center}
		\includegraphics[height=8cm,width=15cm]{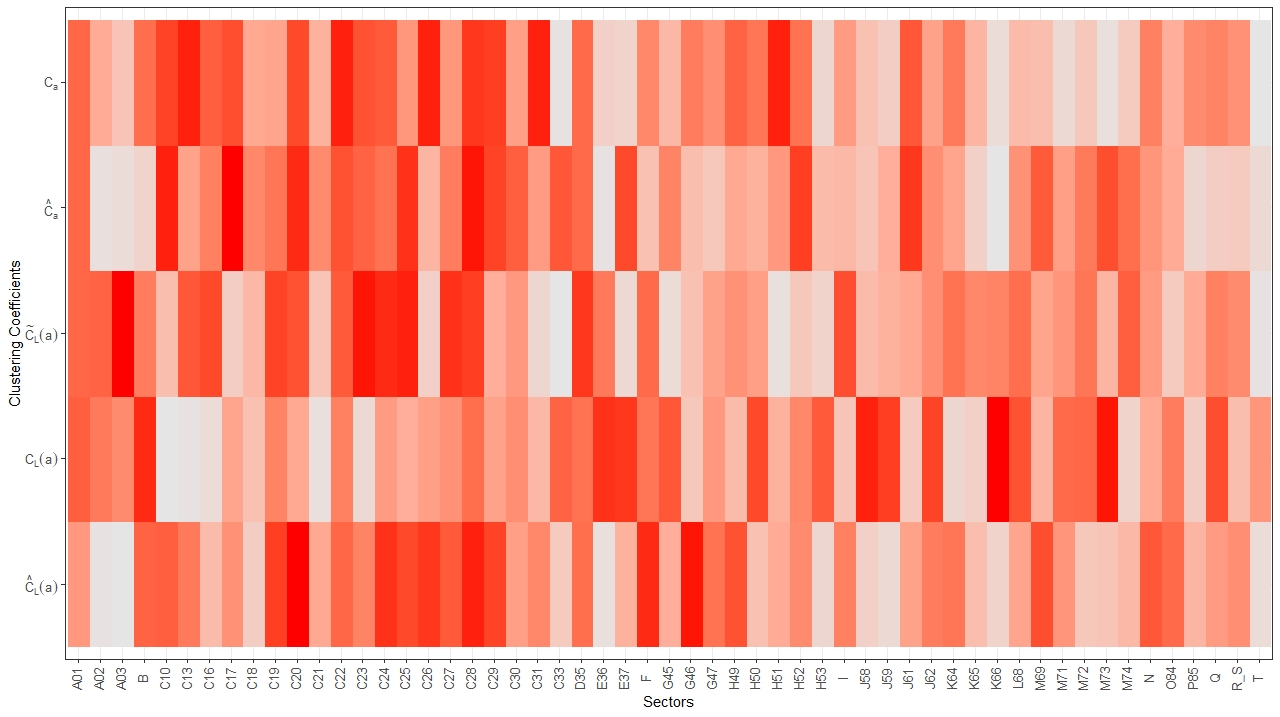}
		\caption{Comparison of rankings of clustering coefficients $\hat{C}_{L}(a)$, $C_{L}(a)$ and $\tilde{C}_{L}(a)$ computed for each sector $a$. The comparison is also extended to the average clustering based on monoplex coefficients. In particular $\hat{C}_{a}$ and $C_{a}$ are the clustering coefficients of each layer based on formulas provided in \cite{Fagiolo2007} and \cite{CleGra} for weighted and directed networks. Darker red means higher ranking.}
		\label{fig:ClustS}
	\end{center}
\end{figure}

In Figure \ref{fig:ClustS}, attention has been also paid to the level of interconnection of sectors. Also in this case, the average multilayer coefficients has been compared with the monoplex versions. We notice again a positive correlation between them but with some significant differences. Focusing on specific sectors, we notice a higher level of interconnections in manufacturing sectors. In particular, sectors related to chemical, mineral, metal products and machinery and equipments (as ``C20", ``C23", ``C25", ``C28") show the highest coefficients. Other relevant sectors are wholetrade (``G46") and constructions (``F").

\section{Conclusions}
\label{Conclusions}
Clustering coefficient has gained increasing attention in network theory. In the literature several proposals have been provided for monoplex networks; however, the study of local and global clustering coefficients for multilayer networks deserves more attention. 

In this paper, we focus on clustering coefficients for weighted and directed multilayer \deleted{node-aligned} networks. The proposed coefficients generalise the alternative proposals provided in the literature for directed monoplex networks and extend existing coefficients for undirected multilayer networks.

The approach has been evaluated using the data taken from the World Input-Output DataBase. Results show the efficiency of the coefficients in capturing the degree of interconnection at different levels, emphasizing the role of countries and sectors in the whole trade taking into account both inter-layer and intra-layer flows.




\bibliographystyle{model5-names}\biboptions{authoryear}

\newpage
\appendix
\section{Lists of sectors and countries} \label{Appendix A}

\begin{table}[H]
	\footnotesize
	\begin{center}
		\begin{tabular}{llllll}
			\hline\hline
			\textbf{Number} & \textbf{Name} & \textbf{Code}
			& 			\textbf{Number} & \textbf{Name} & \textbf{Code}
			\\
			\hline
			1&Australia&AUS &
			2&Austria&AUT \\
			3&Belgium&BEL &
			4&Bulgaria&BGR \\
			5&Brazil&BRA&
			6&Canada&CAN \\
			7&Switzerland&CHE&
			8&China&CHN \\
			9&Cyprus&CYP&
			10&Czech Republic&CZE \\
			11&Germany&DEU&
			12&Denmark&DNK \\
			13&Spain&ESP&
			14&Estonia&EST \\
			15&Finland&FIN&
			16&France&FRA \\
			17&United Kingdom&GBR&
			18&Greece&GRC \\
			19&Croatia&HRV&
			20&Hungary&HUN \\
			21&Indonesia&IDN&
			22&India&IND \\
			23&Ireland&IRL&
			24&Italy&ITA \\
			25&Japan&JPN&
			26&Korea&KOR \\
			27&Lithuania&LTU&
			28&Luxembourg&LUX \\
			29&Latvia&LVA&
			30&Mexico&MEX \\
			31&Malta&MLT&
			32&Netherlands&NLD \\
			33&Norway&NOR&
			34&Poland&POL \\
			35&Portugal&PRT&
			36&Romania&ROU \\
			37&Russian Federation&RUS&
			38&Slovak Republic&SVK \\
			39&Slovenia&SVN&
			40&Sweden&SWE \\
			41&Turkey&TUR&
			42&Taiwan&TWN \\
			43&United States&USA&
			44&Rest of the World&ROW \\
			\hline
	\end{tabular}\end{center}
	\caption  {Lists of 44 countries and areas in the WIOD table.}
	\label{table_WIOD_countries}
\end{table}

\begin{table} [H]
	\centering
	\resizebox{\textwidth}{!}{
		\begin{tabular}{lll}
			\hline\hline
			\textbf{Sector} & \textbf{Description} & \textbf{Code} \\
			\hline
			1&Crop and animal production, hunting and related service activities&A01\tabularnewline
			2&Forestry and logging&A02\tabularnewline
			3&Fishing and aquaculture&A03\tabularnewline
			4&Mining and quarrying&B\tabularnewline
			5&Manufacture of food products, beverages and tobacco products&C10-C12\tabularnewline
			6&Manufacture of textiles, wearing apparel and leather products&C13-C15\tabularnewline
			7&Manufacture of wood and of products of wood and cork, except furniture; manufacture of articles of straw and plaiting materials&C16\tabularnewline
			8&Manufacture of paper and paper products&C17\tabularnewline
			9&Printing and reproduction of recorded media&C18\tabularnewline
			10&Manufacture of coke and refined petroleum products &C19\tabularnewline
			11&Manufacture of chemicals and chemical products &C20\tabularnewline
			12&Manufacture of basic pharmaceutical products and pharmaceutical preparations&C21\tabularnewline
			13&Manufacture of rubber and plastic products&C22\tabularnewline
			14&Manufacture of other non-metallic mineral products&C23\tabularnewline
			15&Manufacture of basic metals&C24\tabularnewline
			16&Manufacture of fabricated metal products, except machinery and equipment&C25\tabularnewline
			17&Manufacture of computer, electronic and optical products&C26\tabularnewline
			18&Manufacture of electrical equipment&C27\tabularnewline
			19&Manufacture of machinery and equipment n.e.c.&C28\tabularnewline
			20&Manufacture of motor vehicles, trailers and semi-trailers&C29\tabularnewline
			21&Manufacture of other transport equipment&C30\tabularnewline
			22&Manufacture of furniture; other manufacturing&C31 C32\tabularnewline
			23&Repair and installation of machinery and equipment&C33\tabularnewline
			24&Electricity, gas, steam and air conditioning supply&D35\tabularnewline
			25&Water collection, treatment and supply&E36\tabularnewline
			26&Sewerage; waste collection, treatment and disposal activities; materials recovery; remediation activities and other waste management services &E37-E39\tabularnewline
			27&Construction&F\tabularnewline
			28&Wholesale and retail trade and repair of motor vehicles and motorcycles&G45\tabularnewline
			29&Wholesale trade, except of motor vehicles and motorcycles&G46\tabularnewline
			30&Retail trade, except of motor vehicles and motorcycles&G47\tabularnewline
			31&Land transport and transport via pipelines&H49\tabularnewline
			32&Water transport&H50\tabularnewline
			33&Air transport&H51\tabularnewline
			34&Warehousing and support activities for transportation&H52\tabularnewline
			35&Postal and courier activities&H53\tabularnewline
			36&Accommodation and food service activities&I\tabularnewline
			37&Publishing activities&J58\tabularnewline
			38&Motion picture, video and television programme production, sound recording and music publishing activities; programming and broadcasting activities&J59 J60\tabularnewline
			39&Telecommunications&J61\tabularnewline
			40&Computer programming, consultancy and related activities; information service activities&J62 J63\tabularnewline
			41&Financial service activities, except insurance and pension funding&K64\tabularnewline
			42&Insurance, reinsurance and pension funding, except compulsory social security&K65\tabularnewline
			43&Activities auxiliary to financial services and insurance activities&K66\tabularnewline
			44&Real estate activities&L68\tabularnewline
			45&Legal and accounting activities; activities of head offices; management consultancy activities&M69 M70\tabularnewline
			46&Architectural and engineering activities; technical testing and analysis&M71\tabularnewline
			47&Scientific research and development&M72\tabularnewline
			48&Advertising and market research&M73\tabularnewline
			49&Other professional, scientific and technical activities; veterinary activities&M74 M75\tabularnewline
			50&Administrative and support service activities&N\tabularnewline
			51&Public administration and defence; compulsory social security&O84\tabularnewline
			52&Education&P85\tabularnewline
			53&Human health and social work activities&Q\tabularnewline
			54&Other service activities&RS\tabularnewline
			55&Activities of households as employers; undifferentiated goods- and services-producing activities of households for own use&T\tabularnewline
			56&Activities of extraterritorial organizations and bodies&U\tabularnewline
			\hline
	\end{tabular}}
	\caption  {Lists of 56  sectors in the WIOD table}
	\label{table_WIOD_industries}
\end{table}







\end{document}